# Sub-Piconewton Force Detection using Micron-Size Wire Deflections

L. Chevry and J.-F. Berret*

*Matière et Systèmes Complexes, UMR 7057 CNRS Université Denis Diderot Paris-VII, Bâtiment Condorcet
10 rue Alice Domon et Léonie Duquet, 75205 Paris (France)*

**Abstract:** The mechanical properties of nanostructured wires obtained by co-assembly of iron oxide particles are studied. The intrinsic magnetic properties of the wires are used to induce and quantify the bending of the one-dimensional objects. From the relationship between the deflection and the magnitude of the magnetic field, the elastic rigidity and Young modulus of the wires are determined. Young moduli in the megapascal range are obtained and are consistent with sub-piconewton force detection.

## 1 - Introduction

Nanowires are anisotropic objects with diameters in the nanometer range and lengths in the micron range. They represent one of the most active research topics of the last 10 years owing to their unique applications in mesoscopic physics and fabrication of nanoscale devices.[1] The one-dimensional growth of nanowires is obtained using well established and controlled techniques, such as oriented crystallization, chemical vapor deposition, epitaxy, template-assisted synthesis and substrate etching.[1,2] In nanoscience, nanowires are foreseen for applications that cover a wide range of phenomena, including sensing,[3-6] actuation,[7-11] manipulation,[12-14] guidance and scaffolding for brain-computer interfaces.[15] In some studies, nanowires are arranged in vertical arrays forming patterned substrates identical to a "Fakir carpet".[16-19] It was discovered that such substrates are compatible with the culture and the growth of living cells, opening new possibilities for mechano-transduction studies and for the delivery of chemicals.[17,20] Recently, vertical nanowire arrays were used as sensors to measure cellular forces of neurons during their migration.[16] The forces were calculated from the displacements of their fluorescently labeled extremities. For semiconductor wires synthesized by crystallization techniques however, the Young moduli $E$ are of the order of 100 GPa,[21] and the sensitivity in the detection of forces remains in the upper piconewton range.[16] The limitations of techniques based on optical microscopy are related to the spatial resolution in the tracking of the wires. To detect forces with an improved resolution and forces of lower magnitudes, one strategy consists in reducing the Young modulus of the wires. This can be realized by modifying the internal structure of the one-dimensional structures.

The strategy developed in the present study relies on a bottom-up co-assembly method using nanoparticles and polymers as building blocks, instead of atoms or molecules. The building blocks are polyelectrolytes and polymer-coated particles and the assembly process uses finely tuned electrostatic interactions. With this technique, wires of diameter 0.1 – 1 µm and lengths comprised between 1 and 200 µm are fabricated.[22] Because the anisotropic objects studied here have diameter above 100 nm, they will be called wires in the following (instead of nanowires). Inside the wires, the particles are held together thanks to the physical cross-linking of the oppositely charged monomers. In terms of structures, the wires described previously bear strong similarities with capsules and thin films fabricated via layer-by-layer assembly.[23] The mechanical properties of layer-by-layer colloids and substrates were studied by atomic force microscopy and revealed values of the Young modulus in the range 0.1 – 10 MPa.[24-27] The elastic rigidity of the membrane was found to be dependent on the nature of the polyelectrolytes, the thickness of the layer and more generally on the physico-chemical conditions, including the pH and the molecular weight.[26, 27] It is important to note here that although electrostatic complexation is a non-covalent binding process, the resulting multilayers exhibit solid-state mechanical properties. Young moduli in the megapascal range correspond to those of soft rubbers and elastomers. The results on the layer-by-layer assemblies also suggest that the electrostatic complexation, if applied to wire structures could generate soft solid-like mechanical properties with high potential benefits in force detection.

In the present paper, we investigate the mechanical responses of nanostructured wires made from magnetic iron oxide particles. Values of the elastic modulus are found to be in the megapascal range, and consistent with highly sensitive detection of forces, typically below one piconewton. With diameters of 0.1 – 1 µm and length comprised between 1 µm and 100 µm, these objects can serve as a functional platform for detecting pico to nano forces.

## 2 - Materials and Methods

*Magnetic wires synthesis*: Wires were formed by electrostatic complexation between oppositely charged nanoparticles and copolymers.[22, 28] The particles were 7.7 nm iron oxide nanocrystals ($\gamma$-$Fe_2O_3$, maghemite) synthesized by polycondensation of metallic salts in alkaline aqueous media.[29] An extensive characterization of the nanometric $\gamma$-$Fe_2O_3$ using various techniques, including vibrating sample magnetometry, dynamic light scattering, zetametry and transmission electron microscopy (TEM, Fig. 1a) is provided in Supporting Information (S1 – S2). To improve their stability, the cationic particles were coated with $M_W = 2100 \, g \, mol^{-1}$ poly(sodium acrylate) (Aldrich) using the precipitation-redispersion process.[30] This process resulted in the adsorption of a highly resilient 3 nm polymer layer surrounding the particles. The copolymer used for the wire synthesis was poly(trimethylammoniumethylacrylate)-*b*-poly(acrylamide) with molecular weights $11000 \, g \, mol^{-1}$ for the charged block and $30000 \, g \, mol^{-1}$ for the neutral block.[31, 32] Fig. 1b displays a TEM image of a wire segment where the individual particles are held tightly together and form a core cylindrical structure. The internal structure of the wires was disclosed by small-angle neutron scattering and revealed a distance between neighboring particle surface of about 1 nm.[31, 33] Fig. 1c displays a transmission optical image of the $\gamma$-$Fe_2O_3$ wires. The wires are polydisperse and their length is characterized by a log-normal distribution (Fig. 1d). For the bending experiments, the wire sample was characterized by an average length $L_0 = 30 \, \mu m$ and a polydispersity of 0.65. For this sample, the lengths were comprised between 1 and 100 µm, few of them being however larger than 100 µm. Studied by scanning electron microscopy, the average diameter $D$ of the wires was estimated at 400 nm.[5] Electrophoretic mobility and $\zeta$–potential measurements made with a Zetasizer Nano ZS

Malvern Instrument showed that the wires were electrically neutral.[28] The shelf life of the co-assembled structures is of the order of several years.

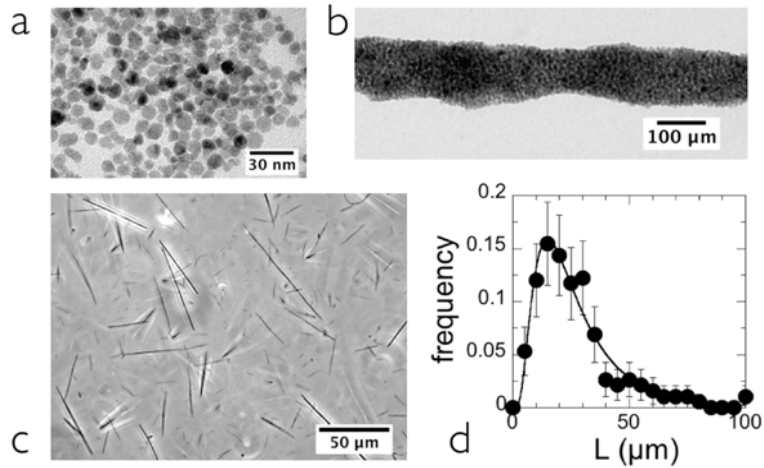

*Figure 1* : a and b) Transmission electron microscopy (TEM) images of 7.7 nm γ-Fe$_2$O$_3$ (maghemite) nanoparticles and of magnetic wires, respectively. c) Phase-contrast image of magnetic wires dispersed in water observed with a 40× objective. d) Length distribution of the wires. The data are adjusted using a log-normal function characterized by an average length of 30 μm and a polydispersity of 0.65. For the bending experiments, wires with lengths higher than 100 μm were isolated and investigated.

*Transmission Electron Microscopy* : TEM on nanomaterials was carried out on a Jeol-100 CX microscope at the SIARE facility of Université Pierre et Marie Curie (Paris 6). It was utilized to characterize the sizes of the γ-Fe$_2$O$_3$ NPs and the magnetic wires (S1, S3).

*Transmission optical microscopy*: Phase-contrast images of the wires were acquired on an IX71 inverted microscope (Olympus) equipped with 40× and 60× objectives. For optical microscopy, 35 μl of a dispersion were deposited on a glass plate and sealed into to a Gene Frame® (Abgene/Advanced Biotech) dual adhesive system. The glass plate was introduced into a homemade device generating both static and rotational magnetic fields, thanks to two pairs of coils working with a 90°-phase shift. An electronic set-up allows measurements on broad ranges of frequency (1 mHz - 100 Hz) and magnetic fields (0 – 20 mT). The image acquisition system consisted of a Photometrics Cascade camera (Roper Scientific) working with Metaview (Universal Imaging Inc.). Images of wires were digitized and treated by the ImageJ software (http://rsbweb.nih.gov/ij/).

*Theoretical background and data analysis:* Submitted to a constant magnetic excitation $\vec{H}$, a nanowire of length $L$, diameter $D$ and magnetization $\vec{m}$ experiences a torque of the form:

$$\vec{\Gamma}_{Mag} = \mu_0 V \vec{m} \wedge \vec{H} \qquad (1)$$

where $\mu_0$ is the vacuum permeability and $V$ the volume ($V = \frac{\pi}{4}D^2 L$) of the wire. For a superparamagnetic wire of susceptibility $\chi$, Eq.1 becomes:[11]

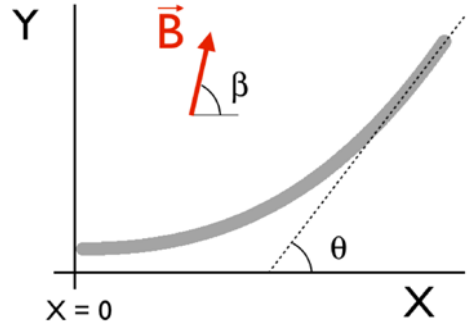

*Figure 2: Schematic representation of a bent nanowire of length L due to the application of a magnetic field. For $x < 0$, the wire is adsorbed on the substrate (part not shown). $\beta$ denotes the angle between the initial wire orientation and that of the applied field B, whereas $\theta$ is related to the orientation of the wire extremity with respect to the horizontal. The profile of bent wires $y(x)$ is analyzed using Cartesian coordinates.*

$$\Gamma_{Mag} = \frac{1}{2}\mu_0 V \Delta\chi H^2 \sin(2\beta) \qquad (2)$$

Here, $\Delta\chi = \chi^2/(2+\chi)$ and $\beta$ is the angle between the wire and the orientation of the applied field (Fig. 2). The quantity $\Delta\chi$ was measured experimentally from the study of the reorientation kinetics of wires submitted to steep changes in the direction of the magnetic field. For wires made of 7.7 nm particles, $\Delta\chi = 0.16$. In the deflection experiment of a beam attached by one of its extremities, the moment $M$ applied to the beam is related to the radius of curvature $R_C$ by:[34]

$$\frac{M}{EI} = \frac{1}{R_C} \qquad (3)$$

where $E$ is the Young modulus and $I$ the second moment of inertia. For a cylindrical beam, $I = \pi D^4/64$. In the bending experiments performed in this work and illustrated in Fig. 2, one part of the wire is physically adsorbed on the substrate, the other one remaining free to move. With increasing magnetic field, $\beta$ being fixed, the wire bends more and more, and the angle $\theta$ between the wire extremity and the horizontal reference approaches that of $\beta$. For large deflections, the angle made by every elementary segment of the wire with respect to the magnetic field varies, and so does the corresponding elementary torque. Replacing $M$ in Eq. 3 by $\Gamma_{Mag}$ from Eq. 2 is hence permitted only for small deflections, *i.e.* for $\sin(2\beta)$ remaining constant along the contour length, or for $\theta$ being close to zero. In such cases, the radius of curvature expresses as:

$$R_C = \frac{\mu_0 E D^2}{8\Delta\chi L \sin(2\beta)} \frac{1}{B^2} \qquad (4)$$

The previous equation shows that the radius of curvature is proportional to the Young modulus and inversely proportional to the magnetic field squared. In Eq. 4, we also assume the equality $B = \mu_0 H$. The Young modulus can then be written as:

$$E = 8\left(\frac{\Delta\chi L}{\mu_0 D^2}\right) R_C B^2 \sin(2\beta) \qquad (5)$$

In the range of validity of Eqs. 4 and 5, it is also expected that the radius of curvature remains constant along the wire length. In Cartesian coordinates, the curvature $1/R_C$ expresses as:

$$1/R_C = y''(x)/(1 + y'(x)^2)^{3/2} \qquad (6)$$

where $y(x)$ denotes the profile of the bent wire. In the analysis, we will use Eqs. 4 – 6 for the estimation of the Young modulus, and we will also test the limit of validity of the different assumptions.

## 3 - Results and discussion

To be able to manipulate single objects and observe them by microscopy, diluted nanowire dispersions were prepared. The iron concentration of these dispersions was of the order of c = 0.001 wt. %, corresponding to 1000 wires per µL. In water, due to gravity, the objects settled down to the bottom of the measuring cell and adsorbed onto the glass slide *via* van der Waals or residual electrostatic interactions. The physical adsorption was a relatively slow process, of the order of the hour and it enabled us to operate individual wires with the external magnetic field.

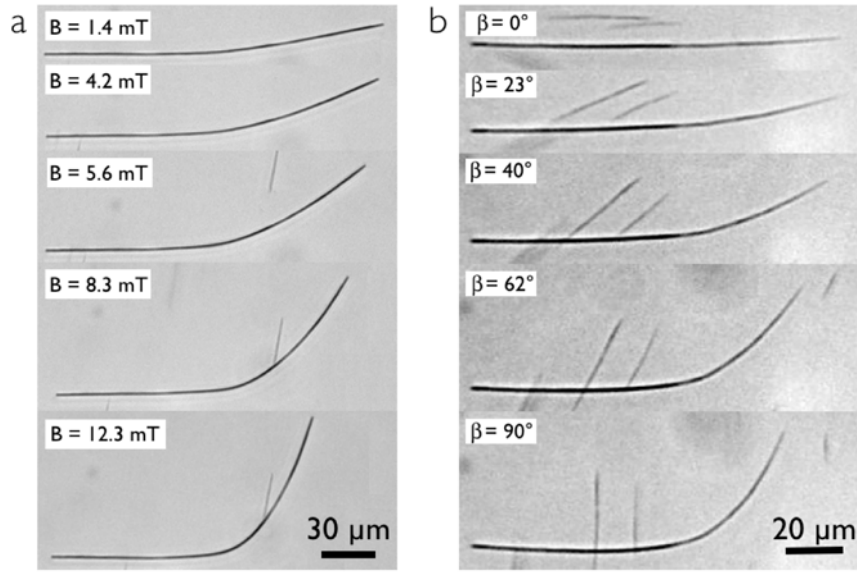

***Figure 3:*** *a) A 190 µm-wire was adsorbed over 65 µm of its length and submitted to a magnetic field oriented at β = 81° with respect to the adhered part. For field increasing between 1.4 and 12.3 mT, the deflection increases continuously. b) a 132 µm-wire was adsorbed over 60 µm of its length and was submitted to a rotating field of intensity B = 7 mT. at the frequency 0.05 Hz. In a) and b), thin wires located above the bottom of the slide are visible and indicate the orientation of the field.*

Partially adsorbed wires were thus characterized by a part that strongly adhered to the glass, and a part that remained free to move. For the bending experiments, wires longer than 100 μm were selected. In the experiments discussed here, it was checked that although the moving part of the wire was close to the surface, it responded to the field in a reproducible way. The magnetic field induced deflection was achieved using two sets of operating conditions. In the first set, the field had a fixed orientation $\beta$ and its amplitude was increased stepwise. In the second experiment, a low-frequency rotating field was applied, the field strength remaining unchanged during the rotation. Fig. 3a and 3b display the deflection of partially adhered wires obtained under the two conditions. In Fig. 3a, a 190 μm-wire was adsorbed over 65 μm of its length and submitted to a magnetic field oriented at $\beta = 81°$ with respect to the adhered part. At low amplitude ($B = 1.4$ mT), the deflection is small and increases with increasing field. In the experiment of Fig. 3b, a 132 μm-wire was adsorbed over 60 μm of its length and was submitted to a rotating field of intensity ($B = 7$ mT). In each, wires located above the bottom of the slide are visible and indicate the orientation of the field. The data in Figs. 3a and 3b are representative of the bending properties studied here.

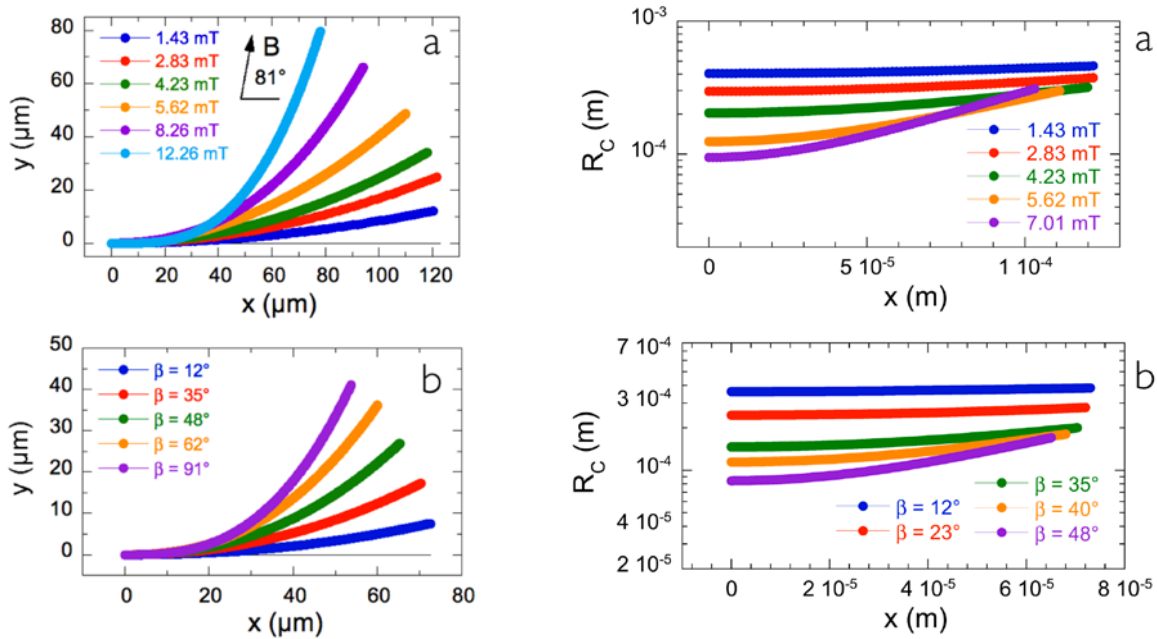

*Figure 4:* Bending profiles $y(x)$ of the wires submitted to the experimental conditions of Fig. 3. a) Magnetic field of fixed direction ($\beta = 81°$) and increasing strength. b) Magnetic field of fixed strength $B = 7$ mT and increasing angle.
*Figure 5*: The radius of curvature $R_C(x)$ was computed from the bending profile of Figs. 4 according to Eq. 6. At low magnetic field or low angle, $R_C(x)$ remains a constant along the contour for the two configurations tested: a) Magnetic field of fixed direction ($\beta = 81°$) and increasing strength. b) Magnetic field of fixed strength $B = 7$ mT and increasing angle.

The data in Figs. 3 were digitalized and the bending profiles expressed in the Cartesian coordinates $y(x)$ were derived. These profiles are shown in Figs. 4a and 4b for various magnetic field strengths and orientation angles. As already mentioned, these two parameters have the strongest impact on the bending. The largest deflections observed are of the order of 60% of the wire length. The $y(x)$-profiles were found to be well accounted for by power laws of the form $y(x) \sim x^\alpha$, with $\alpha \sim 2$ at low deflections, *i.e.* for magnetic field below $6\ mT$ or orientation angles below 45°.

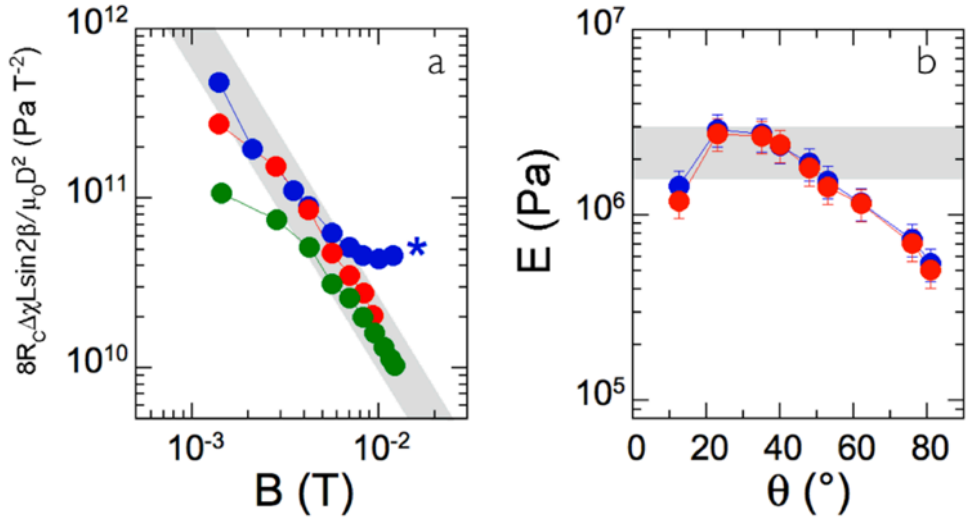

***Figure 6***: *a) The quantity $8R_C \Delta\chi L \sin(2\beta)/\mu_0 D^2$ is shown as a function of the magnetic field for three different configurations: i) $L = 96\ \mu m$ and $\beta = 31°$ (blue symbols); ii) $L = 72\ \mu m$ and $\beta = 61°$ (red symbols), and iii) $L = 125\ \mu m$ and $\beta = 81°$ (green symbols). Here, $R_C$ is the radius of curvature, $\Delta\chi = \chi^2/(2 + \chi)$ where $\chi$ is the magnetic susceptibility, L and D the length and diameter of the wire. The grey area corresponds to the function $E/B^2$, where the Young modulus $E = (1.6 \pm 0.2) \times 10^6\ Pa$. The star associated with the data at $\beta = 31°$ highlights the saturation of the deflection. b) Young modulus E as a function of the angle $\beta$. The grey area corresponds to the range $E = (2.3 \pm 0.7) \times 10^6\ Pa$. The two sets of data points originate from two different analyses. In blue, the $y(x)$-profiles were fitted using a power law dependence with an adjustable exponent. In red the exponent was fixed at 2.*

Above these limits, the exponent increased progressively up to $\alpha = 3$ (S4). The radius of curvature $R_C$ as defined in Eq. 6 was then computed as a function of the abscissa $x$. Figs. 5a and 5b illustrate the $R_C(x)$-variations for the two types of experiments performed, and they disclose that at low magnetic field or low angle, $R_C$ remains constant. Typical values are in the range $10^{-5} - 10^{-3}$ m, *i.e.* of the order of the wire length. As the field intensity or the orientation angle was increased, the radius decreases in the region close to the adherent part ($x \gtrsim 0$). This result is actually expected since this is the region where the deformation and the bending is the strongest. For the estimation of the Young modulus, we will consider the minima of the function $R_C(x)$. In Fig. 6a, the ratio $8R_C \Delta\chi L \sin(2\beta)/\mu_0 D^2$ is plotted as a function of the magnetic field for three different experimental configurations: *i)* $L = 96\ \mu m$ and $\beta = 31°$ (blue symbols); *ii)* $L = 72\ \mu m$ and $\beta = 61°$ (red symbols), and *iii)* $L = 125\ \mu m$ and $\beta = 81°$ (green symbols). According to Eq. 4, the quantity $8R_C \Delta\chi L \sin(2\beta)/\mu_0 D^2$ is expected to vary inversely with the square of the magnetic field at low deflections, the prefactor being the Young modulus $E$. A power law variation with an exponent -2 is indeed observed at low magnetic field, but surprisingly it is also found on the entire $B$-range investigated.

The data from the three experiments agree very well with each other, and least-square calculations yield a Young modulus $E = (1.6 \pm 0.2) \times 10^6$ Pa. The star in Fig. 6a points out the run performed with a field oriented at $\beta = 31°$. In this assay, the deflection and the radius of curvature were shown to saturate with increasing field. This leveling-off behavior corresponds to the situation where the extremity of the wire is collinear to the magnetic field, i.e. $\theta = \beta$ (S5). Above this threshold, the bending profile remains unchanged whatever the field applied. From the data obtained as a function of the angle, Eq. 5 is used and the Young modulus $E = 8\mu_0 \Delta\chi L R_C B^2 \sin(2\beta)/\mu_0 D^2$ is plotted against $\beta$. For magnetic field angle $\beta < 45°$, $E$ is nearly constant $E = (2.3 \pm 0.7) \times 10^6$ Pa, in fair agreement with the previous determination. Values of the order of megapascal are typical of soft rubbers and elastomers, and also typical of capsules and thin films built by layer-by-layer deposition.[24-27] Defined as the distance over which the correlations in the direction of the tangents are lost due to thermal motions, the persistence length $L_P$ of the wires can be estimated through the expression $L_P = EI/k_B T$, where the product $EI$ is the flexural rigidity and $k_B T$ the thermal energy.[34] One finds $L_P = 0.5 \pm 0.1$ m and $L_P = 0.7 \pm 0.2$ m for the experiments of Fig. 3a and 3b respectively, and $EI$ of the order of $3\times 10^{-21}$ Nm$^2$. Since $L_P \gg L$, wires appear as rigid bodies in optical microscopy (Fig. 1c). As compared to biological filaments[35,36] such as actin and microtubule filaments ($L_P \sim 1$ mm, $E = 1$ GPa), the wires studied here have much more persistent. These differences come from geometrical parameters such as the diameter of the objects. Compared to flexible magnetic filaments obtained by association of micrometer beads and studied by Gast et al.[37,38] and by Bibette et al.[39], the Young moduli determined here are higher by a factor nearly 1000.[40] In terms of elastic modulus, the present wires are thus intermediate between biological filaments and the flexible magnetic filaments.

To evaluate the minimum force detectable using such a nano-device, we consider the configuration where a wire is fixed by one extremity, the other end being submitted to a single and localized force. Considering the geometrical and mechanical characteristics of the wires, the force applied at one end will produce a displacement $y(x = L)$ of this extremity. For small displacements, the force expresses as:[34,41]

$$F = \frac{3}{64} \pi E \frac{D^4}{L^3} y(x = L) \qquad (7)$$

Taking for the minimum value of $y(x = L)$ the optical microscopy resolution (typically 1 µm), the minimum forces detected by such wires are $F = 6\times 10^{-14}$ N and $F = 1\times 10^{-12}$ N for 50 µm and 20 µm length wire, respectively. The same orders of magnitude are retrieved using the expression of the magnetic torque $\Gamma_{Mag}$ (Eq. 2) and dividing this torque by the length of the deflected part. By reducing the length down to 10 µm, forces up to 10 pN can be measured from deflected wires. These forces are in the piconewton to the sub-piconewton range, indicating that the wires are appropriate to detect forces that are active at the micron scale and intervene in numerous fields such as microfluidics and cellular biology.[34]

# 4 - Conclusion

We investigated the mechanical properties of nanostructured wires made from the assembly on iron oxide nanoparticles. We exploited the intrinsic magnetic properties of the wires to induce and quantify the bending of these anisotropic objects. The wires put under scrutiny partially adhered onto a substrate, the free moving part being deflected by the application of the field. A wire-bending model valid at low deflections was developed in agreement with the linear response theory[34, 40] and predicted *i)* the constancy of the radius of curvature $R_C$ over the wire contour, and *ii)* a quadratic dependence of the curvature with respect to the field strength. The experimental profiles of bent wires were analyzed and confirmed the $R_C(B) \sim B^{-2}$ behavior of the radius of curvature. At large deflections, $R_C$ was replaced by its minimum value and though, the $B^{-2}$ scaling was still observed. This simple criterion, associated with a straightforward data analysis can be used to retrieve the values of the Young modulus with an improved accuracy. These values were found to be in the megapascal range, demonstrating that such non-covalent structures do exhibit solid-state mechanical properties. From the Young modulus estimates, the persistence length of the wires was determined, and found of the order of 1 m, *i.e.* 1000 longer than their length. Such high values of persistence length agree with the observation that wires appear as rigid bodies under a microscope. For magnetic fields larger than 10 mT, reversible deflections up to 60% were observed. At a nanometer level, the nanoparticle network enclosed in the polyelectrolyte matrix is thus able to deform considerably. However, it does not show signs of plastic flow or rupture. This resilience of the wire structure to repetitive deflections is illustrated in the SI Section (S6). Reversible responses to high mechanical solicitations make these systems attractive for applications. Elastic rigidity in the megapascal range are finally consistent with highly sensitive detection of forces, typically below one picoNewton. With diameters of 0.1 – 1 μm and length comprised between 1 μm and 100 μm, these objects serve as a functional platform for detecting pico to nano forces.

## Acknowledgments

Discussions with A. Cebers and J. Fresnais are acknowledged. The Laboratoire Physico-chimie des Electrolytes, Colloïdes et Sciences Analytiques (UMR Université Pierre et Marie Curie-CNRS n° 7612) is acknowledged for providing us with the magnetic nanoparticles.

## Supporting Information

The Supporting Information section S1 provides the characterization of the nanoparticle in terms of sizes and distribution. S2 deals with the vibrating sample magnetometry results obtained of iron oxide nanoparticle dispersions. The desalting transition towards the fabrication of nanostructured wires is described in S3, whereas S4 and S5 show data on the bending profiles of the wires submitted to various magnetic field conditions. This information is available free of charge *via* the Internet at xxx.

## References

1. Y. N. Xia, P. D. Yang, Y. G. Sun, Y. Y. Wu, B. Mayers, B. Gates, Y. D. Yin, F. Kim and Y. Q. Yan, *Advanced Materials*, 2003, **15**, 353-389.
2. S. Xu, Y. Qin, C. Xu, Y. G. Wei, R. S. Yang and Z. L. Wang, *Nature Nanotechnology*, 2010, **5**, 366-373.
3. J. Lee, P. Hernandez, J. Lee, A. O. Govorov and N. A. Kotov, *Nature Materials*, 2007, **6**, 291-295.
4. W. S. Chang, J. W. Ha, L. S. Slaughter and S. Link, *Proceedings of the National Academy of Sciences of the United States of America*, 2010, **107**, 2781-2786.
5. R. Colin, M. Yan, L. Chevry, J.-F. Berret and B. Abou, *Europhysics Letters*, 2012, **97**.
6. Y. Coffinier, S. Szunerits, H. Drobecq, O. Melnyk and R. Boukherroub, *Nanoscale*, 2012, **4**, 231-238.

7. N. Cappallo, C. Lapointe, D. H. Reich and R. L. Leheny, *Physical Review E*, 2007, **76**, 6.
8. B. A. Evans, A. R. Shields, R. L. Carroll, S. Washburn, M. R. Falvo and R. Superfine, *Nano Lett.*, 2007, **7**, 1428 - 1434.
9. U. Chippada, B. Yurke, P. C. Georges and N. A. Langrana, *Journal of Biomechanical Engineering-Transactions of the Asme*, 2009, **131**.
10. J. V. I. Timonen, C. Johans, K. Kontturi, A. Walther, O. Ikkala and R. H. A. Ras, *Acs Applied Materials & Interfaces*, 2010, **2**, 2226-2230.
11. B. Frka-Petesic, K. Erglis, J.-F. Berret, A. Cebers, V. Dupuis, J. Fresnais, O. Sandre and R. Perzynski, *Journal of Magnetism and Magnetic Materials*, 2011, **323**, 1309-1313.
12. P. Dhar, Y. Y. Cao, T. M. Fischer and J. A. Zasadzinski, *Physical Review Letters*, 2010, **104**.
13. A. Hultgren, M. Tanase, E. J. Felton, K. Bhadriraju, A. K. Salem, C. S. Chen and D. H. Reich, *Biotechnology Progress*, 2005, **21**, 509 - 515.
14. M. M. Song, W. J. Song, H. Bi, J. Wang, W. L. Wu, J. Sun and M. Yu, *Biomaterials*, 2010, **31**, 1509-1517.
15. F. Johansson, M. Jonsson, K. Alm and M. Kanje, *Experimental Cell Research*, 2010, **316**, 688-694.
16. W. Hällström, M. Lexholm, D. B. Suyatin, G. Hammarin, D. Hessman, L. Samuelson, L. Montelius, M. Kanje and C. N. Prinz, *Nano Letters*, 2010, **10**, 782-787.
17. W. Hällström, T. Mårtensson, C. Prinz, P. Gustavsson, L. Montelius, L. Samuelson and M. Kanje, *Nano Letters*, 2007, **7**, 2960-2965.
18. G. Piret, E. Galopin, Y. Coffinier, R. Boukherroub, D. Legrand and C. Slomianny, *Soft Matter*, 2011, **7**, 8642-8649.
19. W. Kim, J. K. Ng, M. E. Kunitake, B. R. Conklin and P. Yang, *Journal of the American Chemical Society*, 2007, **129**, 7228-7229.
20. A. K. Shalek, J. T. Robinson, E. S. Karp, J. S. Lee, D. R. Ahn, M. H. Yoon, A. Sutton, M. Jorgolli, R. S. Gertner, T. S. Gujral, G. MacBeath, E. G. Yang and H. Park, *Proceedings of the National Academy of Sciences of the United States of America*, 2010, **107**, 1870-1875.
21. D. Hessman, M. Lexholm, K. A. Dick, S. Ghatnekar-Nilsson and L. Samuelson, *Small*, 2007, **3**, 1699-1702.
22. J. Fresnais, J.-F. Berret, B. Frka-Petesic, O. Sandre and R. Perzynski, *Advanced Materials*, 2008, **20**, 3877-3881.
23. G. Decher, *Science*, 1997, **277**, 1232 - 1237.
24. L. Richert, A. J. Engler, D. E. Discher and C. Picart, *Biomacromolecules*, 2004, **5**, 1908-1916.
25. N. Delorme, M. Dubois, S. Garnier, A. Laschewsky, R. Weinkamer, T. Zemb and A. Fery, *Journal of Physical Chemistry B*, 2006, **110**, 1752-1758.
26. M. Schonhoff, V. Ball, A. R. Bausch, C. Dejugnat, N. Delorme, K. Glinel, R. V. Klitzing and R. Steitz, *Colloids and Surfaces a-Physicochemical and Engineering Aspects*, 2007, **303**, 14-29.
27. T. Boudou, T. Crouzier, R. Auzely-Velty, K. Glinel and C. Picart, *Langmuir*, 2009, **25**, 13809-13819.
28. M. Yan, J. Fresnais, S. Sekar, J. P. Chapel and J.-F. Berret, *Acs Applied Materials & Interfaces*, 2011, **3**, 1049-1054.

29. R. Massart, E. Dubois, V. Cabuil and E. Hasmonay, *J. Magn. Magn. Mat.*, 1995, **149**, 1 - 5.
30. J.-F. Berret, O. Sandre and A. Mauger, *Langmuir*, 2007, **23**, 2993-2999.
31. J.-F. Berret, A. Sehgal, M. Morvan, O. Sandre, A. Vacher and M. Airiau, *Journal of Colloid and Interface Science*, 2006, **303**, 315-318.
32. J.-F. Berret, *Macromolecules*, 2007, **40**, 4260-4266.
33. J.-F. Berret, *Advances in Colloid and Interface Science*, 2011, **167**, 38-48.
34. J. Howard, *Mechanics of Motor Proteins ans teh Cytoskeleton*, Sinauer Associates, Inc. Publishers, Sunderland, Massachusetts, 2001.
35. P. Venier, A. C. Maggs, M. F. Carlier and D. Pantaloni, *Journal of Biological Chemistry*, 1994, **269**, 13353-13360.
36. J. H. Shin, L. Mahadevan, P. T. So and P. Matsudaira, *Journal of Molecular Biology*, 2004, **337**, 255-261.
37. E. M. Furst and A. P. Gast, *Physical Review E*, 2000, **61**, 6732.
38. S. L. Biswal and A. P. Gast, *Physical Review E*, 2003, **68**, 021402.
39. C. Goubault, P. Jop, M. Fermigier, J. Baudry, E. Bertrand and J. Bibette, *Physical Review Letters*, 2003, **91**, 260802.
40. A. Cebers, *Current Opinion in Colloid & Interface Science*, 2005, **10**, 167-175.
41. J. le Digabel, N. Biais, J. Fresnais, J.-F. Berret, P. Hersen and B. Ladoux, *Lab on a Chip*, 2011, **11**, 2630-2636.

**TOC figure**

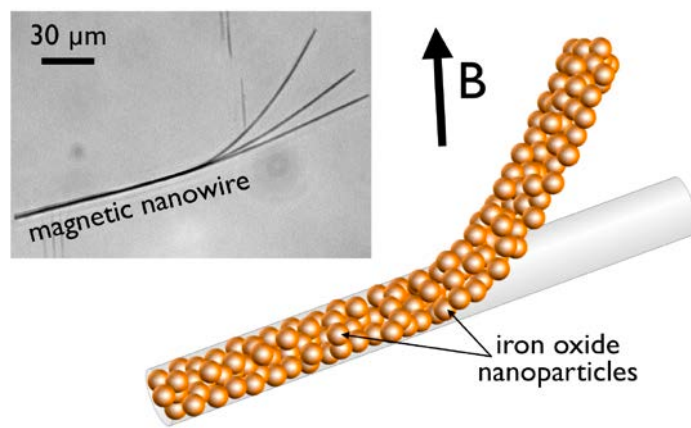

**TOC text**
Micron-size wires obtained by co-assembly of iron oxide particles detect deflection forces in the picoNewton range.